\documentclass[twocolumn,aps]{revtex4}

\usepackage{graphicx} % Include figure commands

\begin{document}

\title{Towards practical classical processing for the surface code: timing analysis}

\author{Austin G. Fowler, Adam C. Whiteside, Lloyd C. L. Hollenberg}
\affiliation{Centre for Quantum Computation and Communication
Technology, School of Physics, The University of Melbourne, Victoria
3010, Australia}

\date{\today}

\begin{abstract}
Topological quantum error correction codes have high thresholds and are well suited to physical implementation. The minimum weight perfect matching algorithm can be used to efficiently handle errors in such codes. We perform a timing analysis of our current implementation of the minimum weight perfect matching algorithm. Our implementation performs the classical processing associated with an $n\times n$ lattice of qubits realizing a square surface code storing a single logical qubit of information in a fault-tolerant manner. We empirically demonstrate that our implementation requires only $O(n^2)$ average time per round of error correction for code distances ranging from 4 to 512 and a range of depolarizing error rates. We also describe tests we have performed to verify that it always obtains a true minimum weight perfect matching.
\end{abstract}

\maketitle

\section{introduction}

Quantum computers promise efficient factoring \cite{Shor94b}, efficient simulation of quantum systems \cite{Lloy96}, and the efficient solution of many other classically intractable problems \cite{Jord10}. The primary barrier to the realization of a quantum computer is the physical realization of quantum gates with sufficiently low error to enable quantum error correction to be used. Topological quantum error correction (TQEC) codes can tolerate error rates of order 1\% \cite{Wang11,Fowl11b} and require only 2-D nearest neighbor interactions, both physically reasonable targets, however the classical processing associated with the error correction is highly nontrivial. Without significant future effort, the classical processing will almost certainly limit the speed of any quantum computer, particularly one with intrinsically fast quantum gates.

In this work, we present a timing analysis of our software performing the classical processing associated with TQEC. This software is by orders of magnitude the fastest currently available. We will review the necessary aspects of the surface code \cite{Brav98,Denn02}, fault-tolerant schemes built on the surface code \cite{Raus07,Raus07d,Fowl12f}, and our classical processing algorithm \cite{Fowl11b} as required. Our goal is to analyze in detail the performance and correctness of our implementation of this algorithm. This implementation is contained in a library match.c and called by our tool Autotune \cite{Fowl12d}, which is designed to prepare a graph problem tailored to arbitrary hardware running a surface code family TQEC scheme.

The discussion is organized as follows. In Section~\ref{Overview}, the basic structure and functionality of our software is described. The library match.c, which performs minimum weight perfect matching \cite{Edmo65a,Edmo65b} is described in more detail in Section~\ref{Matching}. Two versions are discussed. An example of the faster version of the algorithm in action is provided in Section~\ref{example}. The probability of logical errors in the surface code as a function of the physical error rate $p$ is discussed in Section~\ref{Logical errors}. Formatted timing data is presented in Section~\ref{Timing}. Complete raw timing data can be found in the Supplementary Material. Section~\ref{Conclusion} summarizes and points to further work.

\section{Overview}
\label{Overview}

\begin{figure}
\begin{center}
\resizebox{85mm}{!}{\includegraphics{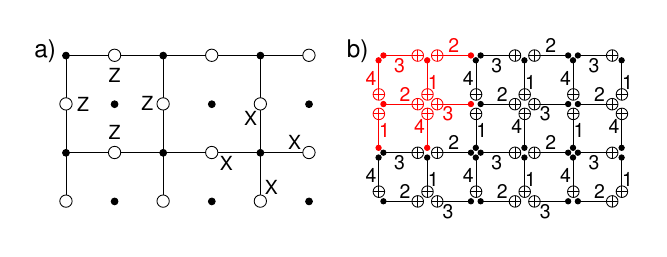}}
\end{center}
\caption{(Color online) a) 2-D lattice of data qubits (circles) and syndrome qubits
(dots) and examples of the data qubit stabilizers. b) Sequence of
CNOTs permitting simultaneous measurement of all stabilizers.
Numbers indicate the relative timing of gates. The highlighted gates
can be tiled to fill the plane.}\label{transversal_sequence}
\end{figure}

Our simulation suite of software is designed to handle arbitrary hardware with arbitrary stochastic error models, however we shall focus on a simple 2-D square lattice of qubits and a standard depolarizing channel for each quantum gate for the purposes of benchmarking and demonstrating correctness. Specifically, we shall study the case of no initialization surface code error detection \cite{Wang11}. A small section of the 2-D array of data and syndrome qubits of the surface code and the required cyclic sequence of CNOTs to simultaneously measure all stabilizers \cite{Gott97} is shown in Fig.~\ref{transversal_sequence}. At the end of each cycle, all syndrome qubits are measured in the $X$ or $Z$ basis according to whether they are being used to measure $X$ or $Z$ stabilizers, respectively.

\begin{figure}
\begin{center}
\resizebox{80mm}{!}{\includegraphics[viewport=100 55 800 525, clip=true]{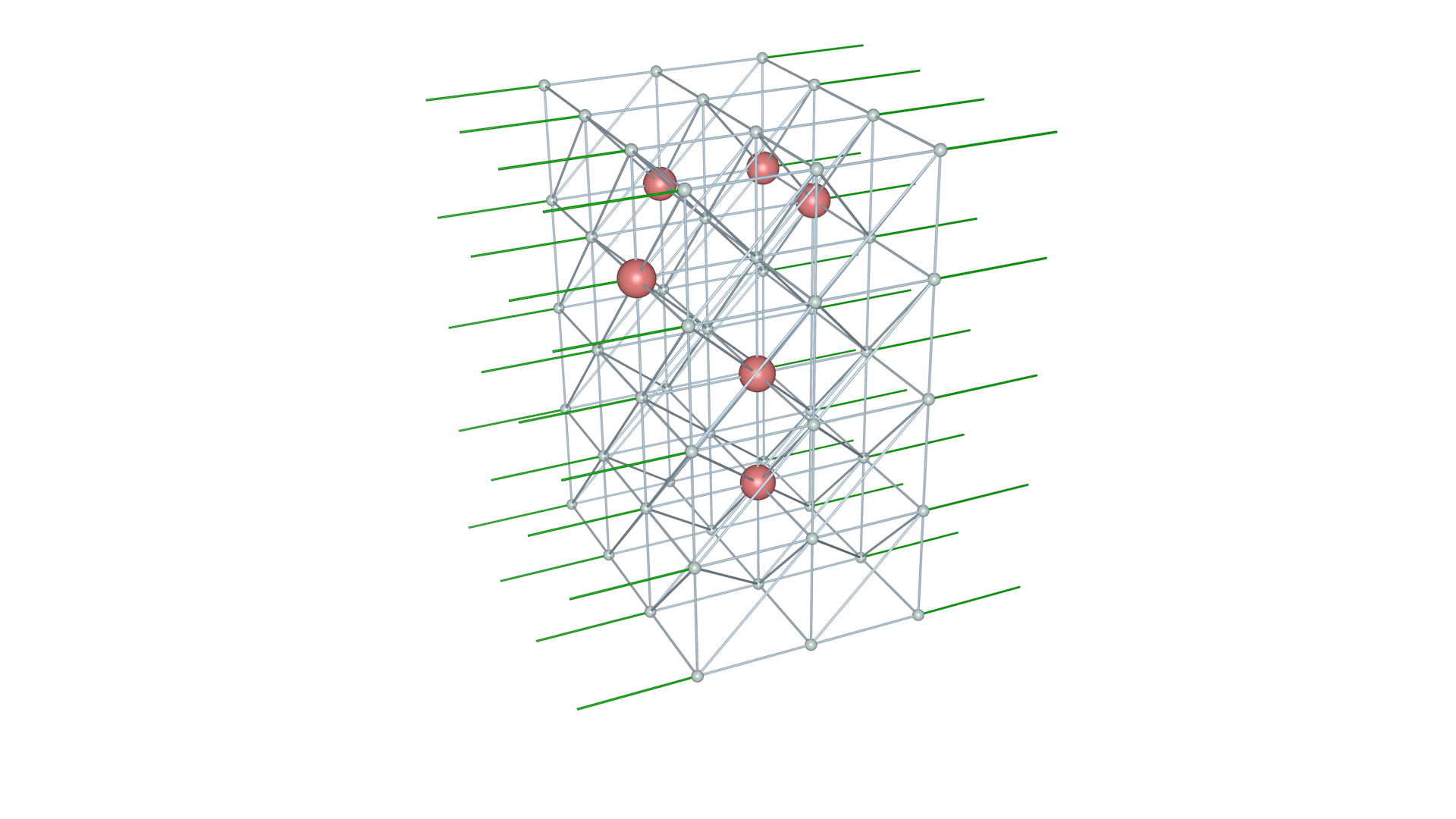}}
\end{center}
\caption{(Color online) Distance 4 example of a lattice of dots and lines with stochastically generated vertices. The distance of a surface code if the length in lines of the shortest topologically nontrivial path, in this case any path connecting opposing boundaries. Dots (small balls) correspond to space-time locations where the endpoints of error chains could potentially be detected. Vertices (large balls) correspond to space-time locations where error chain end points have been detected. Light cylinders link pairs of dots where a pair of vertices could be generated by a single error. Dark cylinders link spatial boundaries to a single dot where a single vertex could be generated by a single error.}\label{problem}
\end{figure}

Random Pauli errors are generated and propagated using a Pauli frame. When errors lead to syndrome measurement value changes, graph vertices are generated at these space-time locations. By pre-analyzing all possible single error processes \cite{Wang11,Fowl12d}, an underlying lattice of dots and lines is also prepared with dots at every location a vertex could potentially be generated and lines between every pair of locations that could have vertices generated by a single error. The first order probability $p_{\rm line}$ of each line is calculated and a weight $w = -\ln(p_{\rm line})$ stored in each line. This is done so that a large positive weight is associated with any line of low probability, ensuring that an algorithm matching vertices impairs using paths of lines with minimum total weight will tend to avoid using low probability lines. Furthermore, a multiple line path will have a weight related to the product of probabilities of its constituent lines. A lattice of dots and lines and stochastically generated vertices (from surface code simulation) is shown in Fig.~\ref{problem}.

In many ways, a lattice plus vertices can be considered an implicit complete graph with an edge between any pair of vertices having weight equal to the minimum weight path between those vertices. The task is to match all vertices in pairs or to neighboring boundaries such that the total weight of all match paths is minimal. The basic algorithm that efficiently solves this problem given a standard graph is the minimum weight perfect matching algorithm \cite{Edmo65a,Edmo65b}. We have extended this algorithm to include the concept of boundaries and permit new vertices to be dynamically added to the graph.

\section{Matching}
\label{Matching}

\begin{figure}
\begin{center}
\resizebox{80mm}{!}{\includegraphics[viewport=100 200 800 450, clip=true]{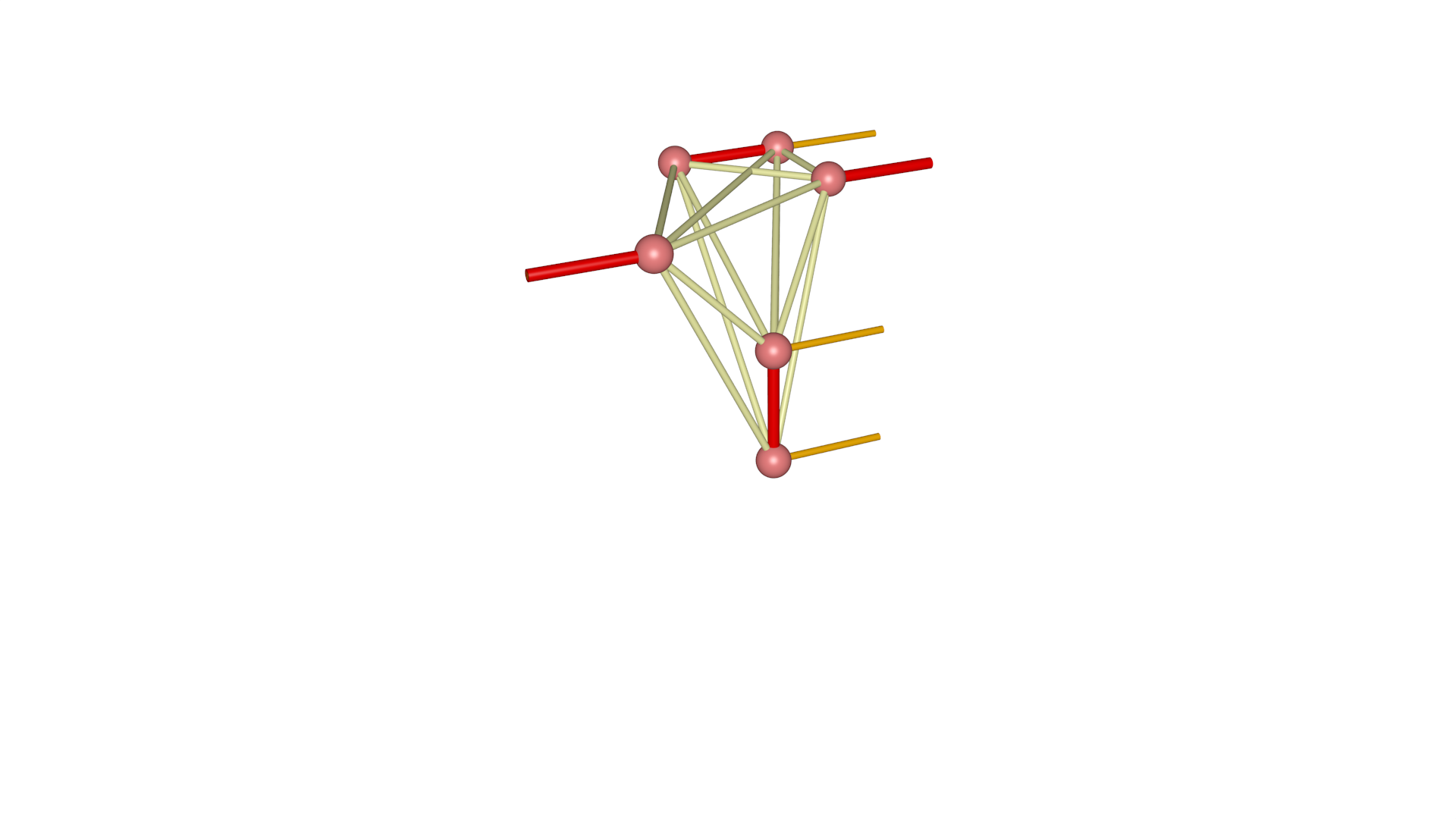}}
\end{center}
\caption{(Color) Output of cmatch when given Fig.~\ref{problem} as input. The underlying lattice is used to construct a complete graph and then discarded. Edges in the matching are shown in red.}\label{complete}
\end{figure}

\begin{figure}
\begin{center}
\resizebox{80mm}{!}{\includegraphics[viewport=100 0 800 525, clip=true]{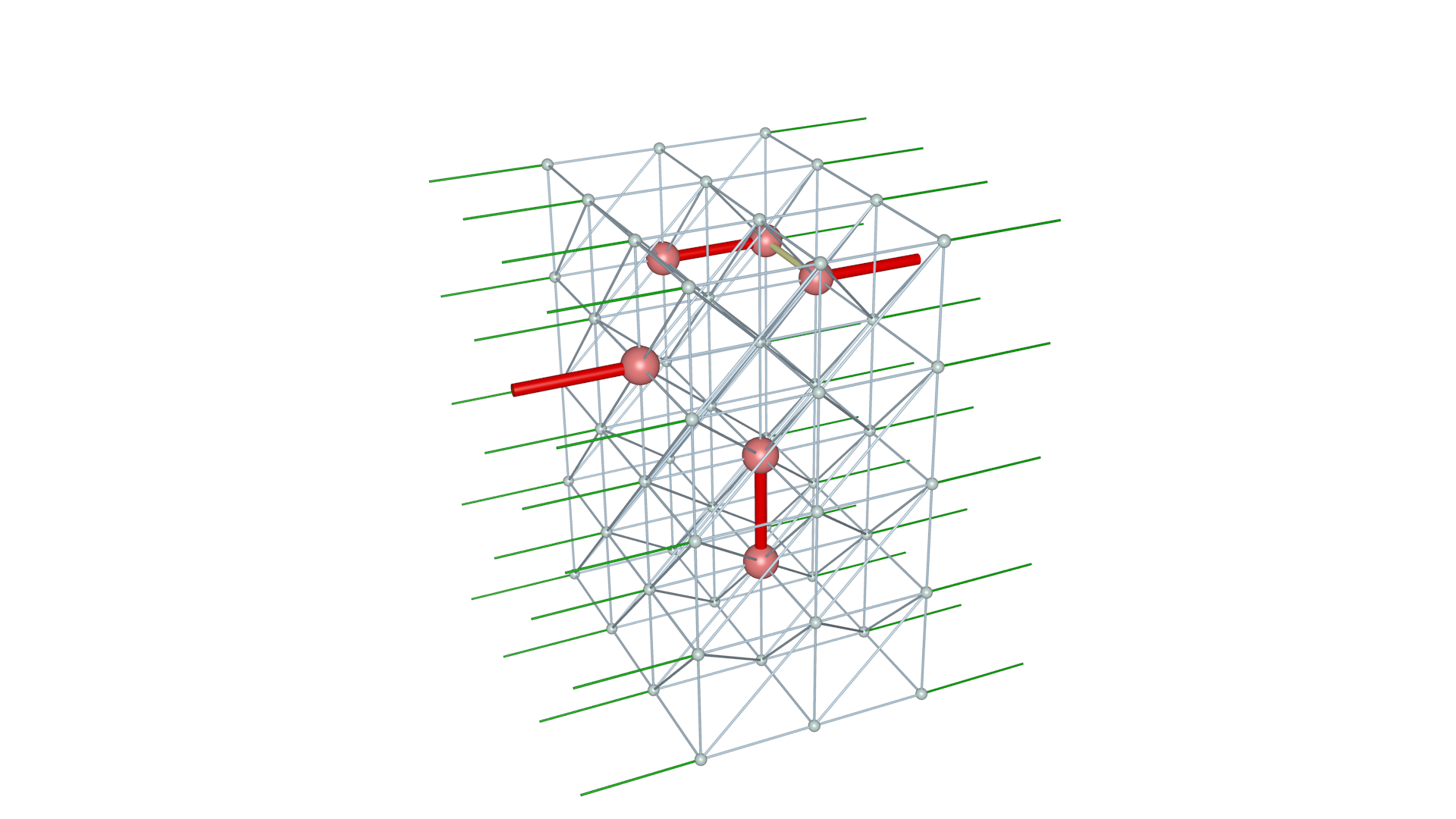}}
\end{center}
\caption{(Color) Output of eodmatch when given Fig.~\ref{problem} as input. Note that only one edge other than those ultimately included in the matching has been created (just visible between the top right two vertices).}\label{eod}
\end{figure}

We have two operational versions of extended minimum weight perfect matching --- complete match \cite{Wang11} which firstly constructs explicit edges between all pairs of vertices no more than approximately $d$ rounds of error correction apart, and edges on demand match \cite{Fowl11b} which only constructs a small number of local edges and adds further edges to the problem as required. The graphs and matchings generated by complete match (cmatch) and edges on demand match (eodmatch) given Fig.~\ref{problem} as input are shown in Fig.~\ref{complete} and Fig.~\ref{eod} respectively. The total weight of matched edges in both cases is identical and in this case the matchings themselves are identical. We have tested cmatch and eodmatch on millions of varied problems, large and small, and always observed identical total weights, strongly implying both implementations are correct.

Cmatch obtains the true minimum weight perfect matching despite only including edges between vertices separated by a finite number of rounds. Vertices separated by a very large number of rounds are always cheaper to match to their nearest boundaries than to one another. By using the weights of the lines in the lattice, we calculate the minimum span of rounds to connect with edges to guarantee a minimum weight matching. Eodmatch also obtains a true minimum weight perfect matching as any required edge will eventually be included during execution.

We now describe the eodmatch algorithm. Some definitions are required. Let $G$ be a graph with vertices $\{v_i\}$,
edges $\{e_{ij}\}$, and edge weights $\{w_{ij}\}$. The graphs we use in eodmatch are implicitly complete, with the weight of an edge between any given pair of vertices defined to be the weight of a minimum weight path between those vertices, and the weight of any edge connecting a vertex to a nearby boundary defined similarly. As such, we shall describe the algorithm as though we have a complete graph. The process of dynamically adding the required edges is just a technical detail.

\begin{figure}
\begin{center}
\resizebox{85mm}{!}{\includegraphics{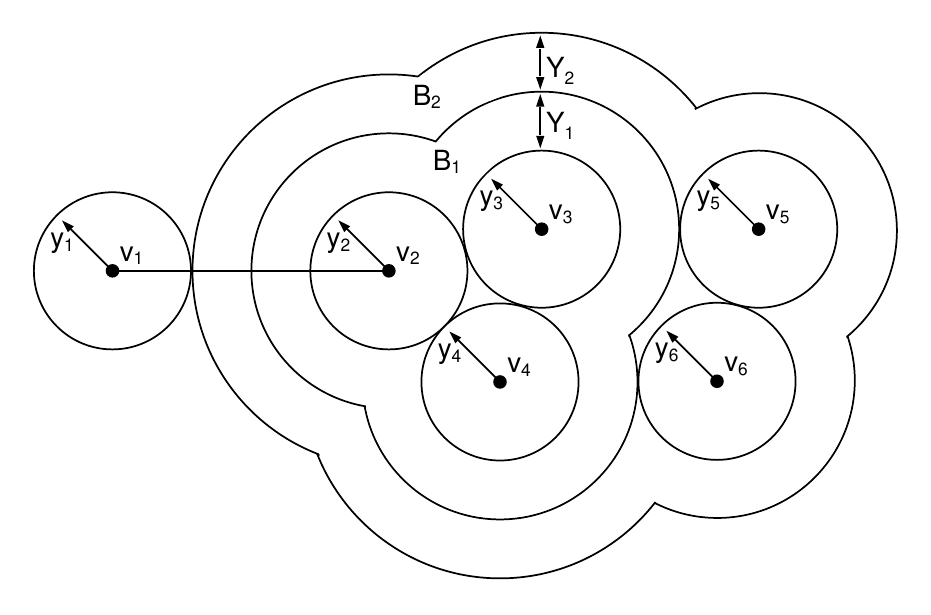}}
\end{center}
\caption{An example of a tight edge. Edge $e_{12}$ has
the property that $w_{12}-y_1-y_2-Y_1-Y_2=0$.}\label{tight_edge}
\end{figure}

Associate with
each vertex $v_i$ a variable $y_i$, which can be thought of as the
radius of a ball centered at $v_i$. Odd sets of vertices can also be
made into blossoms $B_k$ that have their own variables $Y_k$, which
can be thought of as the width of shell around every object in
$B_k$. If a pair of blossoms intersect, then one is contained in the
other. Define an edge $e_{ij}$ to be tight if $w_{ij}-y_i-y_j-\sum
Y_k = 0$, where the sum is over $k$ such that exclusively $v_i$ or
$v_j$ is in $B_k$. This condition is pictorially depicted in
Fig.~\ref{tight_edge}.

\begin{figure}
\begin{center}
\resizebox{85mm}{!}{\includegraphics{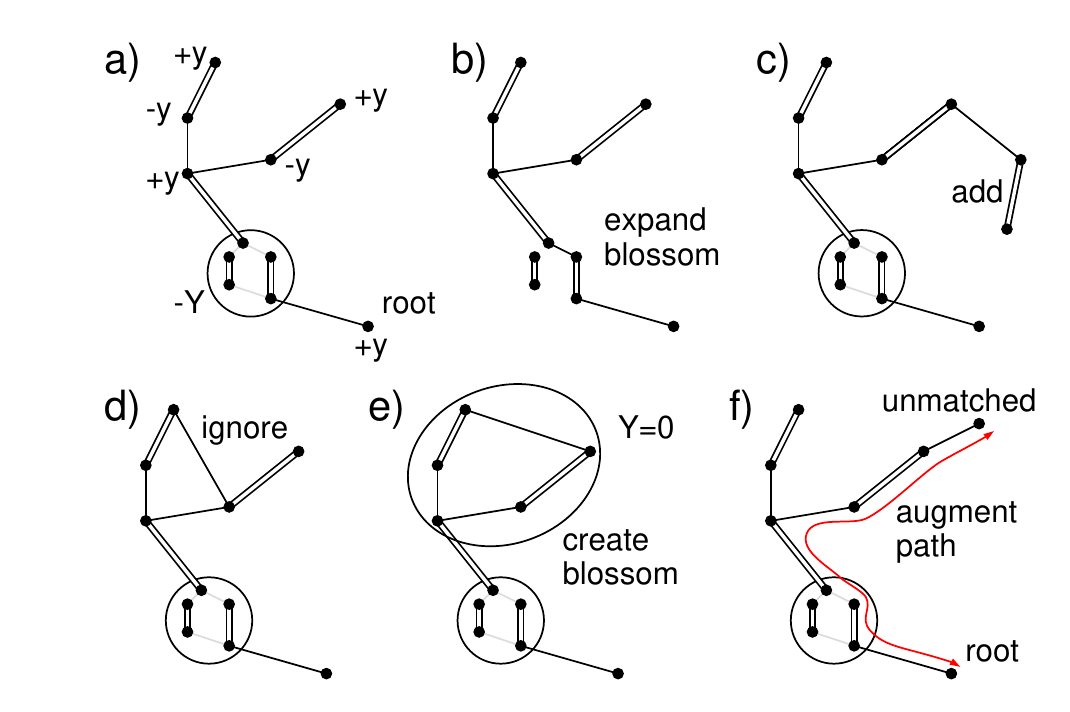}}
\end{center}
\caption{(Color online) All required alternating tree manipulations.
a)~Increase outer node and decrease inner node $y$ values (or $Y$ if the node is a blossom), which will maintain the tightness of all tree edges and potentially creating new tight edges connected to at least one outer node. b)~Inner blossoms
with $Y=0$ can be expanded into multiple inner and outer nodes and potentially some nodes that are no longer part of the tree. c)~Outer--matched tight edges can be used to grow the alternating tree.
d)~Outer--inner tight edges can be ignored as they never grow tighter. e)~Outer--outer tight edges
make cycles that can be used to form blossoms. f)~When another unmatched vertex $v$ is found, or an edge to a boundary $b$, the path from the unmatched vertex within the root node through the alternating tree to $v$ or $b$ is augmented, meaning matched edges become unmatched and unmatched edges become matched. This strictly increases the total number of matched vertices.}\label{Tree}
\end{figure}

Define a node to be a vertex or blossom. Allow edges to possess a label matched or unmatched. Define a blossom to be
unmatched if it contains a vertex not incident on a matched edge. An alternating tree is
a tree of nodes rooted on an unmatched node such that every path
of edges from the root node to a leaf node consists of alternating unmatched and
matched edges. Alternating trees can only branch from the root and
every second node from the root. Define branching nodes to be outer.
Define all other nodes in the alternating tree to be inner.
Fig.~\ref{Tree} shows all necessary alternating tree manipulations.

Given a weighted graph $G$, the following algorithm finds a minimum
weight perfect matching.
\begin{enumerate}
\item If there are no unmatched vertices, return the list of matched edges.

\item Choose an unmatched vertex $v$ to be the root of an alternating tree.

\item If no edges emanating from the outer nodes of the
alternating tree are tight, henceforth called $O$-tight edges,
increase the value of $y$ or $Y$ associated with each outer node
while simultaneously decreasing the value of $y$ or $Y$ associated
with each inner node until an edge becomes $O$-tight, or an inner
blossom node $Y$ variable becomes 0 (Fig.~\ref{Tree}a).

\item If an inner blossom node $Y$ variable becomes 0
and there are still no $O$-tight edges, expand that blossom and
return to 3 (Fig.~\ref{Tree}b).

\item Choose an $O$-tight edge $e$.

\item If $e$ leads to a matched node not already in the
alternating tree, add the relevant unmatched and matched edge and
associated nodes to the alternating tree and return to 3
(Fig.~\ref{Tree}c).

\item If $e$ leads to an inner node, mark $e$ so it is
not considered again during the growth of this alternating tree and
return to 3 (Fig.~\ref{Tree}d).

\item If $e$ leads to an outer node, add the unmatched
edge to the alternating tree. There will now be a cycle of odd
length. Collapse this cycle into a new blossom and associate a new
variable $Y=0$ (Fig.~\ref{Tree}e). Return to 3.

\item If $e$ leads to an unmatched vertex or boundary,
add $e$ to the alternating tree and augment the path
(unmatched$\leftrightarrow$matched) from the unmatched vertex within the root node to the end of $e$
(Fig.~\ref{Tree}f). Destroy the alternating tree, keeping any newly
formed blossoms. Return to 1.
\end{enumerate}

On average, the algorithm only needs to consider a small local
region around each vertex to find another unmatched vertex to pair
with. This is a property of the graphs associated with topological
QEC only, as the probability of needing to consider an edge of
length $l$ decreases exponentially with $l$. This ensures that the
runtime is $O(n^2)$, and that the algorithm can be parallelized to achieve $O(1)$ processing per round.

\section{Eodmatch example}
\label{example}

\begin{figure*}
\begin{center}
\resizebox{140mm}{!}{\includegraphics{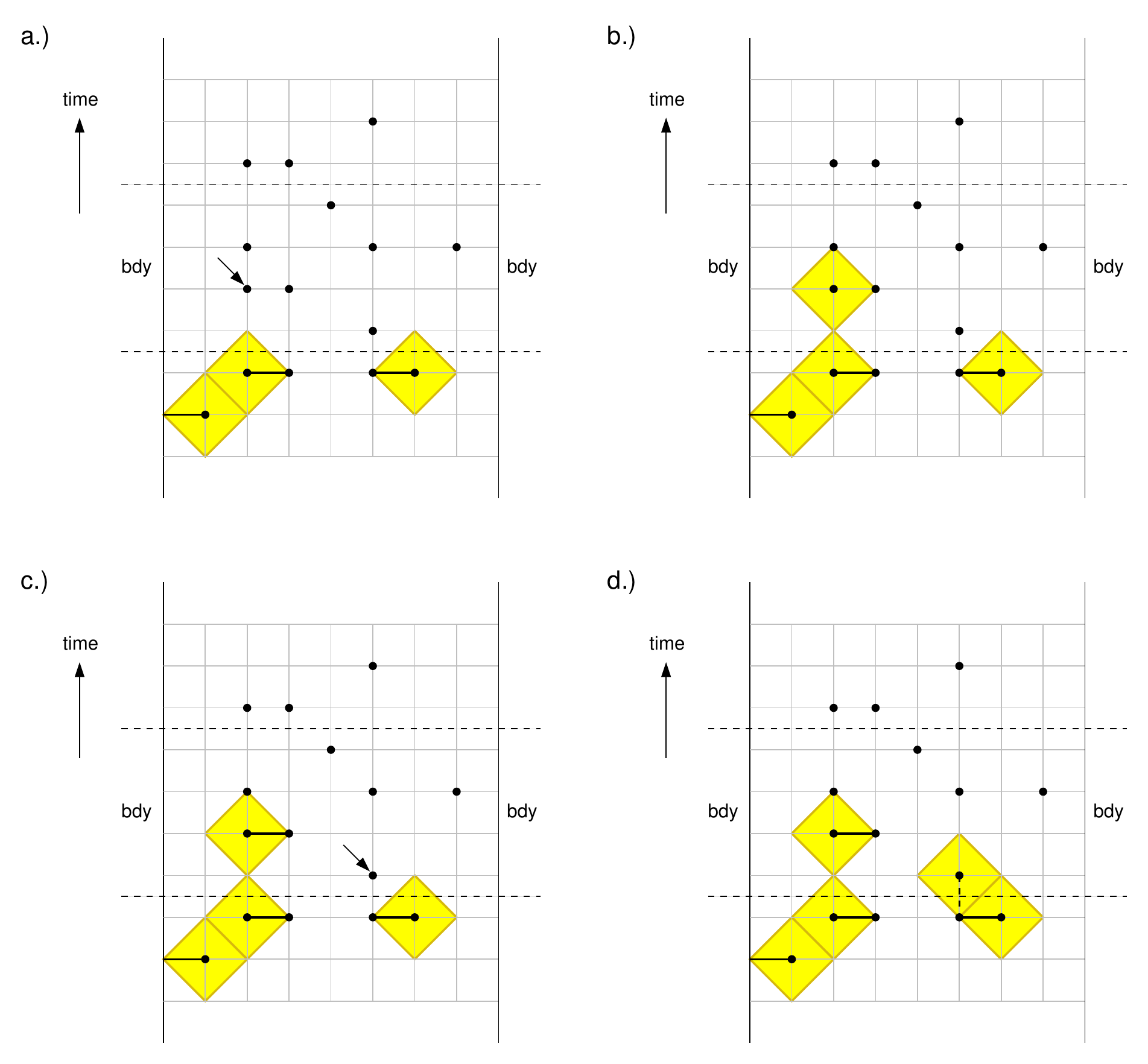}}
\end{center}
\caption{(Color online) a.) Choose an unmatched vertex. b.) Expand exploratory region until other objects encountered. c.) Unmatched vertices encountered, choose one to match to. Choose another unmatched vertex. d.) Expand exploratory region until other objects encountered. Build alternating tree.}\label{ex1}
\end{figure*}

The rules of the previous section are far from intuitive. Let's consider a simple 1-D chain of qubits suffering errors and generating vertices in space and time. Let's assume the underlying lattice is square. Fig.~\ref{ex1}a shows a possible current state of the matching algorithm, with matched vertices far in the past and unmatched vertices in the present and recent past. The goal is to match as many vertices as possible in the active region (between the horizontal dashed lines) without using any data that is too new. The window that defines the active region rolls forward as additional vertices are generated by the quantum computer. The first vertex chosen for matching is indicated with an arrow. It does not matter which vertex is chosen in the active region, however our algorithm has a preference for vertices further in the past.

Fig.~\ref{ex1}b shows a shaded exploratory region around the chosen vertex. This is constructed by performing a breadth first search through the lattice local to the vertex. When any other object is encountered, whether it be a boundary, another exploratory region or another vertex, expansion is halted. In this case, two unmatched vertices and one exploratory region simultaneously terminate expansion. One of these vertices is chosen to be matched to as shown in Fig.~\ref{ex1}c. It does not matter which vertex is chosen, both are valid choices that would lead to a minimum weight perfect matching being obtained. The next vertex is chosen.

In Fig.~\ref{ex1}d, when exploration around the chosen vertex terminates, a matched vertex is encountered and no unmatched vertices. This necessitates the construction of an alternating tree. An alternating tree is a tree with alternating unmatched and matched edges. Alternating trees are only allowed to branch at the root and every second node from the root. Branching nodes are called outer nodes, non-branching nodes are called inner nodes. The alternating tree constructed in Fig.~\ref{ex1}d consists of three nodes, all of which are simple vertices. We will encounter more complex alternating trees shortly.

\begin{figure*}
\begin{center}
\resizebox{140mm}{!}{\includegraphics{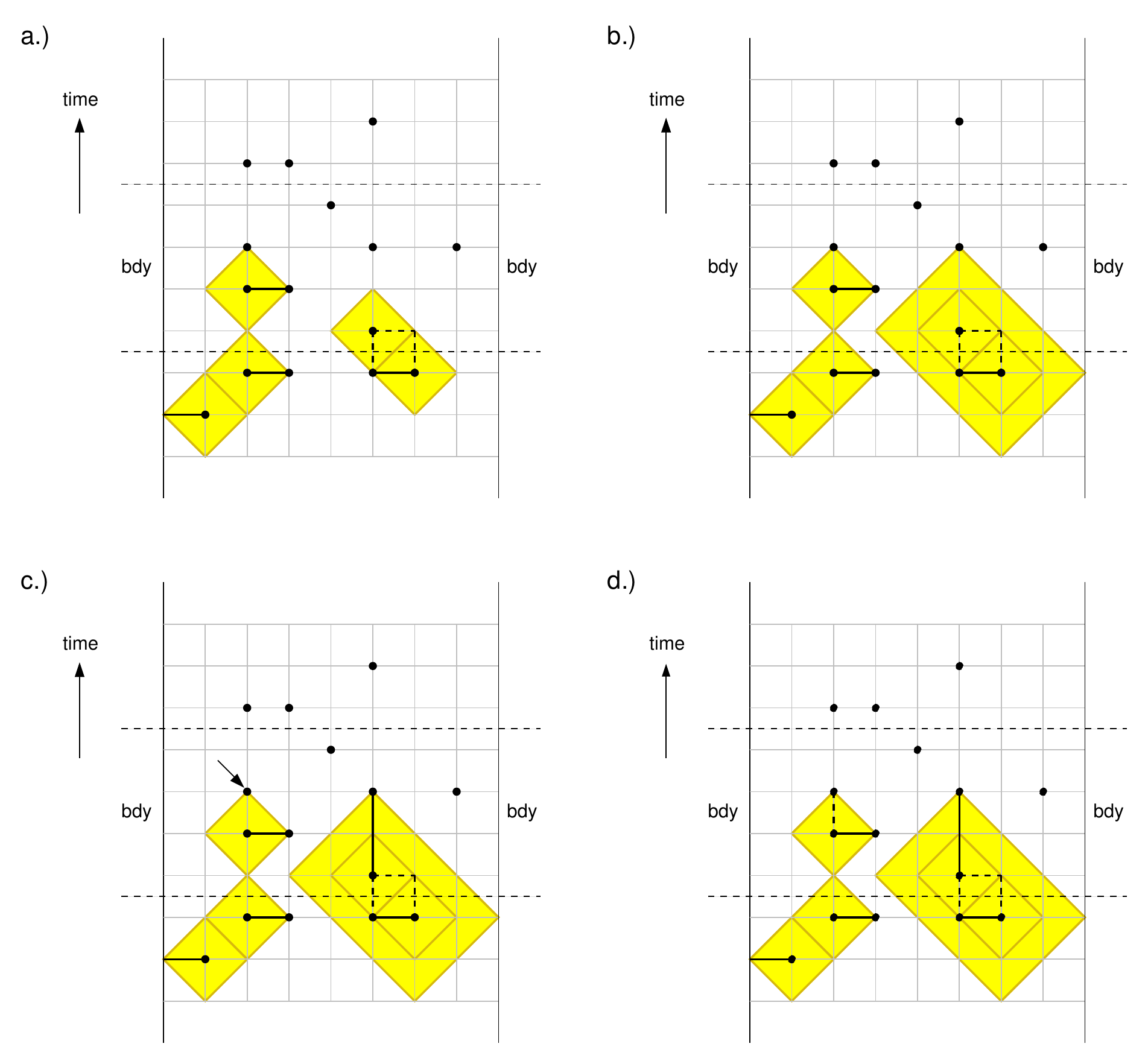}}
\end{center}
\caption{(Color online) a.) Form blossom. b.) Expand exploratory region around blossom until other objects encountered. c.) Match to unmatched vertex. Choose another unmatched vertex. d.) Form alternating tree.}\label{ex2}
\end{figure*}

Our algorithm attempts to expand the exploratory regions around each outer node and contract the exploratory regions around each inner node. This is impossible in this case as the two outer nodes are touching. Instead, a cycle is formed as shown in Fig.~\ref{ex2}a. This cycle is collapsed to form a blossom, leaving an alternating tree with a single outer node that is a blossom containing three vertices.

The exploratory region around the sole outer node in the alternating tree is expanded until other objects are encountered (Fig.~\ref{ex2}b). An unmatched vertex and a boundary are encountered. Two options are available. We could match the edge from the original root vertex to the vertex below it, unmatch the existing matched edge, and then match the resultant unmatched vertex to the nearby boundary. Alternatively, we can match the original root vertex to the newly encountered unmatched vertex. Since this is simpler, we choose this option, the execution of which is shown in Fig.~\ref{ex2}c. The next unmatched vertex chosen is indicated by an arrow. In this case, no expansion of the exploratory region around the vertex is possible. One must instead immediately form an alternating tree consisting of three vertices (Fig.~\ref{ex2}d).

\begin{figure*}
\begin{center}
\resizebox{140mm}{!}{\includegraphics{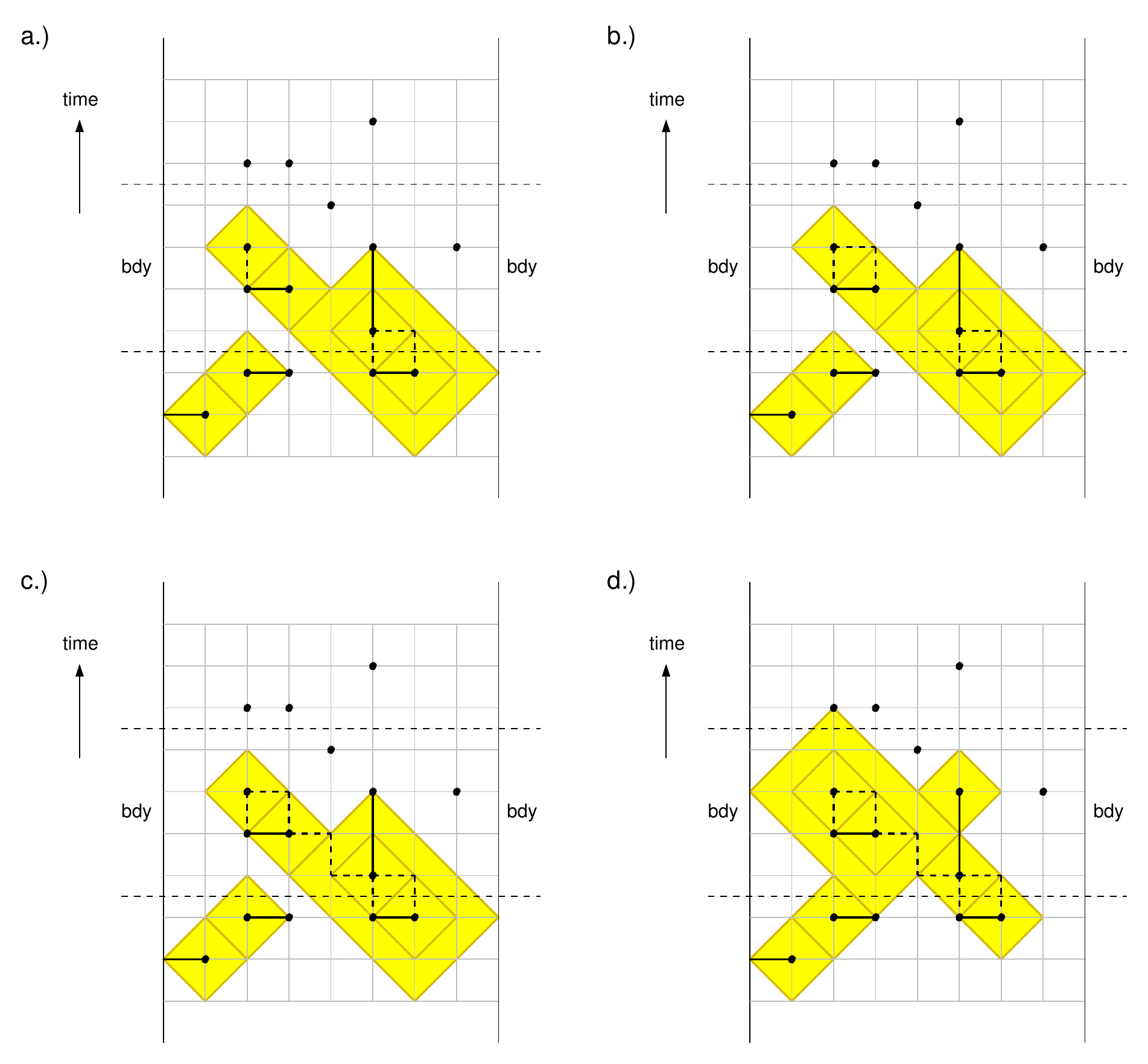}}
\end{center}
\caption{(Color online) a.) Expand outer node exploratory regions, contract inner node exploratory region. b.) Form blossom. c.) Form alternating tree. d.) Expand outer node exploratory regions, contract inner node exploratory region. Forbidden region entered, reverse algorithm execution back to Fig.~\ref{ex2}c. Wait for additional data.}\label{ex3}
\end{figure*}

The outer node exploratory regions are expanded while the inner node exploratory region is contracted (Fig.~\ref{ex3}a). This results in the outer node exploratory regions touching, forming a cycle and thus a blossom (Fig.~\ref{ex3}b). This collapses the alternating tree to a single outer node consisting of a blossom containing three vertices.

The exploratory region around the single blossom outer node cannot be expanded, necessitating the creation of another alternating tree a blossom outer node, then a blossom inner node, then a vertex outer node (Fig.~\ref{ex3}c). The two outer exploratory regions can be expanded while the blossom inner node exploratory region is contracted, however this leads to exploration outside the active region (Fig.~\ref{ex3}d). When this happens, we run our algorithm backwards to the beginning of the current matching attempt, which in this case is Fig.~\ref{ex2}c.

This example hopefully gives a flavor of the algorithm. The salient features we wish to convey to the reader are the algorithm's space-time locality and continuous processing nature. These features enable one to understand the parallelization of the algorithm. We shall explain this by analogy.

Imagine a box being filled with sand using a 2-D array of tubes. Each tube represents a processor. Imagine the rate of sand coming out of each tube represents the difficulty of the matching problem locally. A slower rate of flow implies higher local difficulty. The sand itself represents vertices that have been matched. The rate of flow of all tubes is set below the maximum possible -- pauses are inserted in the algorithm such that it is possible for a tube to be run at greater than the standard rate should it be required. When the problem is locally hard, the rate of flow decreases and a hollow forms locally. When the difficulty of the problem returns to normal, which it must do on average, the rate of flow is increased above the standard rate to fill in this hollow. When the hollow is filled, the rate of flow is brought below the maximum possible again. Local difficulty does not result in global slowdown. Furthermore, surrounding tubes can assist in filling in the hollow. This simple picture explains how one can obtain a globally optimal solution of an infinite size problem in constant average time per round of processing, which is optimal.

Two other techniques for correcting errors in surface codes are being investigated, renormalization \cite{Ducl09} and metropolis \cite{Woot12}. However, neither approach has been successfully applied to the realistic fault-tolerant case. Indeed, in the latest work of the authors of the renormalization approach, minimum weight perfect matching has been used to handle the fault-tolerant case \cite{Brav12}. We are not hopeful that any technique other than matching can be comparably fast and effective in the fault-tolerant case.

\section{Logical errors}
\label{Logical errors}

Strong evidence of the correctness of eodmatch comes from studying the probability of logical error per round of error correction ($p_L$) at depolarizing probabilities $p$ well below threshold. We calculate $p_L$ by simulating $t_{\rm check}$ rounds of faulty quantum computer operation, then turning off errors, capping the matching problem with a perfect round of error correction, applying corrections, checking whether we have an odd or even number of errors along one of the boundaries and recording whether this is different to the previous time we checked. The perfect round of error correction is then undone and another $t_{\rm check}$ faulty rounds simulated and the process repeated.

It may seem that the ideal value of $t_{\rm check}$ is 1 to ensure that no logical errors are missed, however this is not the case. We have observed that many combinations of errors lead to the observation of a logical error if a perfect round of error correction is inserted halfway through it, but no logical error if the perfect round of correction is sufficiently distant. With frequent checking this can mean a benign pattern of errors is counted as several logical errors. Instead, we typically use a value of $t_{\rm check}$ such that a change in the parity of the number of errors observed along a boundary occurs approximately 10\% of the time. We have empirically found that this leads to a logical error rate estimate robust to wide variations of $t_{\rm check}$ about this value. The probability of a change per check is equal to the probability of an odd number of logical errors in $t_{\rm check}$ rounds enabling $p_L$ to be easily calculated.

\begin{figure*}
\begin{center}
\resizebox{140mm}{!}{\includegraphics[viewport=0 55 650 400, clip=true]{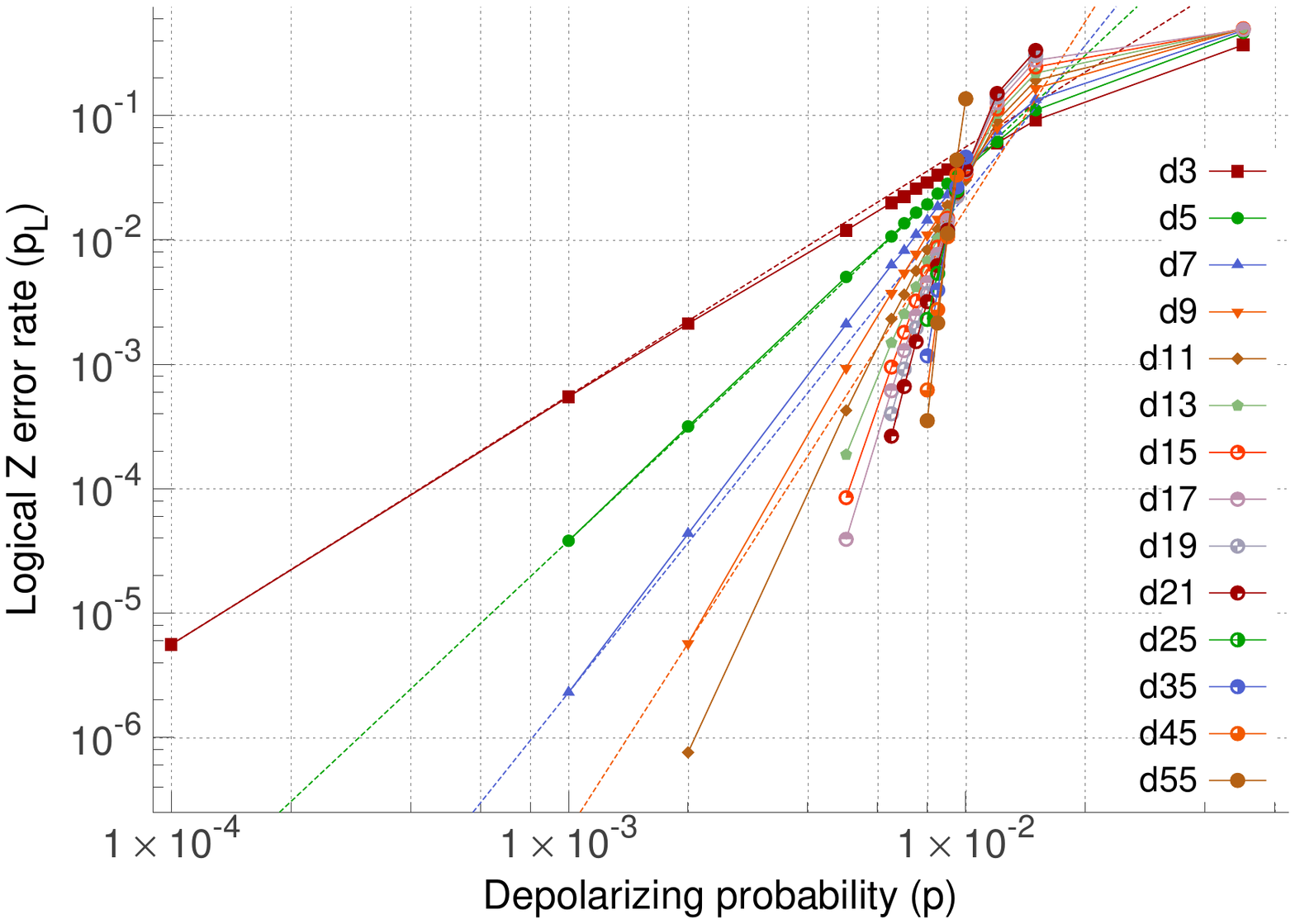}}
\end{center}
\caption{(Color online) Logical $Z$ error rate per round of error correction for surface code distances $d$ and depolarizing noise probabilities $p$. Dashed lines indicate expected low $p$ asymptotic curves for $d=3$, 5, 7 and 9.}\label{logicalz}
\end{figure*}

\begin{figure*}
\begin{center}
\resizebox{140mm}{!}{\includegraphics[viewport=0 55 650 400, clip=true]{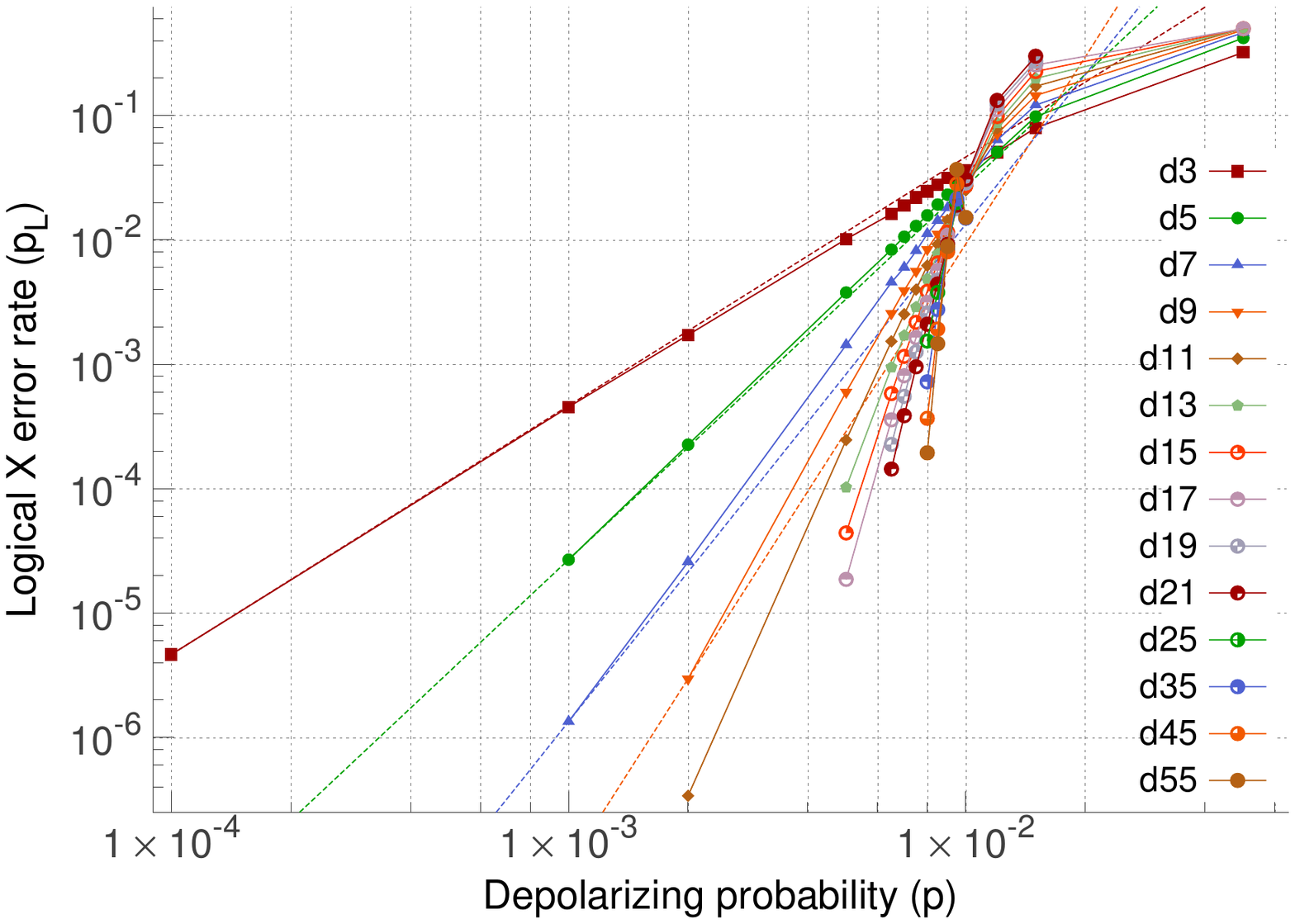}}
\end{center}
\caption{(Color online) Logical $X$ error rate per round of error correction for surface code distances $d$ and depolarizing noise probabilities $p$. Dashed lines indicate expected low $p$ asymptotic curves for $d=3$, 5, 7 and 9.}\label{logicalx}
\end{figure*}

A distance $d$ code can reliably correct $\lfloor (d-1)/2 \rfloor$ errors. At low error rates $p$, clusters of errors are rare and well separated. The probability of suffering a logical error inducing cluster of $n_d=\lfloor (d+1)/2 \rfloor$ errors should therefore be $O(p^{n_d})$ if the full distance of the code is being realized. Figs.~\ref{logicalz}--\ref{logicalx} show the complete set of data we have collected for the square surface code. Polynomials $A_d p^{n_d}$ are drawn through the lowest data point we were able to obtain for distances 3, 5, 7 and 9.

It is computationally expensive to obtain statistics at very low error rates and high distances as very few logical state changes are observed. It is also computationally expensive to obtain data at high error rates and high distances as the minimum weight perfect matching problem becomes more difficult around and above the threshold error rate (0.9\% \cite{Fowl11b}). The raw data used to generate Figs.~\ref{logicalz}--\ref{logicalx}, including timing information, can be found in the Supplementary Material.

The distance 3 and 5 dashed asymptotic curves in Figs.~\ref{logicalz}--\ref{logicalx} agree very well with the data. For higher distances, it is not currently possible to simulate a sufficiently large number of rounds of error correction to obtain sufficient information at low enough probabilities to achieve such tight agreement. Note that the high distance data curves approach the asymptotic curves with a steeper gradient, implying the surface code is capable of regularly correcting temporal clusters of errors containing more errors than the maximum guaranteed to be correctable. This is a generic feature of topological quantum error correction, as a large cluster of errors widely scattered across the code is not dangerous provided the cluster poorly resembles a topologically nontrivial chain of errors connecting distinct boundaries.

\section{Timing}
\label{Timing}

\begin{figure*}
\begin{center}
\resizebox{140mm}{!}{\includegraphics[viewport=0 60 650 430, clip=true]{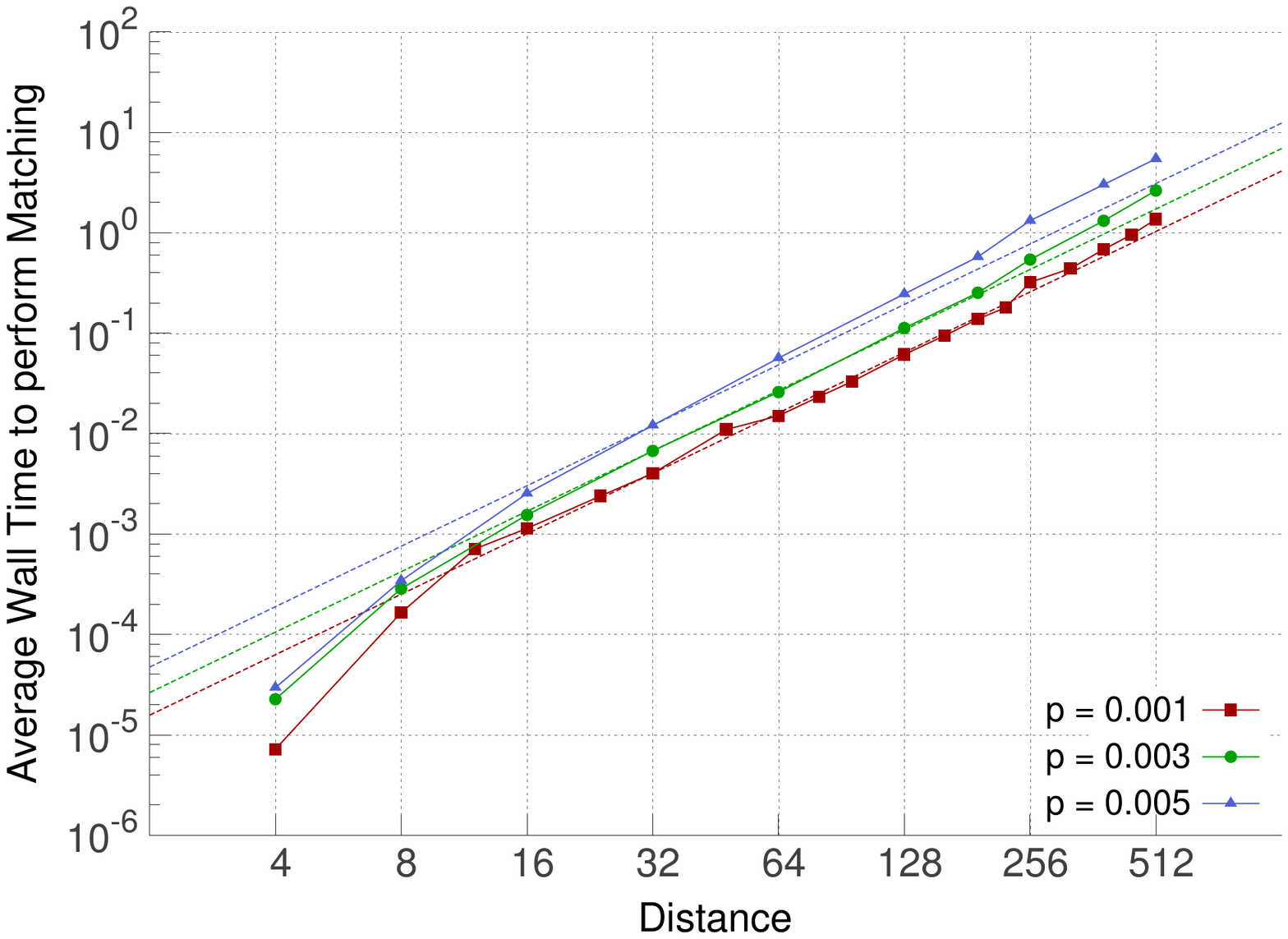}}
\end{center}
\caption{(Color online) Amount time in seconds devoted to each round of matching when simulating a distance $d$ single logical qubit square surface code for depolarizing error rates $p$. Quadratic curves have been included for reference.}\label{timing}
\end{figure*}

The timing information in the Supplementary Material includes everything --- initial bootup of the simulation, the simulation of the underlying quantum computer, problem generation, matching, perfect rounds of error correction to enable logical state change detection, and maintenance of an appropriate Pauli frame. Figs.~\ref{timing} shows the amount of time devoted to each round of matching alone at three different error rates for distances $d=4$, 8, 16, $\ldots$, 512. The quadratic scaling of required time with distance is well demonstrated. At small $d$ nearby boundaries prevent the growth of large blossoms leading to increased performance. At very high $d$ memory access effects lead to a slight slowdown. Note that real computer systems are too complex to provide perfectly smooth graphs of time scaling even with long time averaging as the interplay of different levels of cache and RAM leads to measurable deviations from the ideal scaling.

\begin{figure*}
\begin{center}
\resizebox{140mm}{!}{\includegraphics[viewport=0 60 650 430, clip=true]{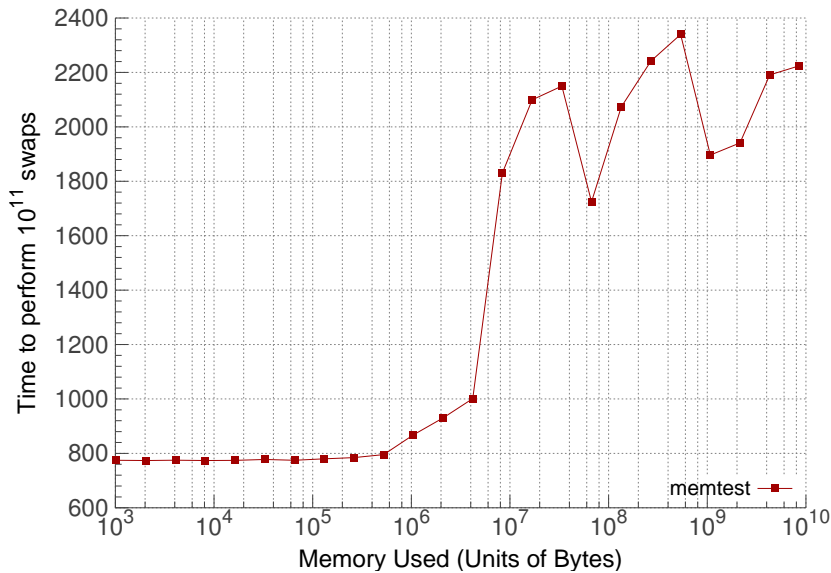}}
\end{center}
\caption{(Color online) Average time in seconds required to perform $10^{11}$ swaps of randomly chosen pairs of integers in arrays of increasing size.}\label{swap}
\end{figure*}

To illustrate the complexity of modern computer memory systems, we have generated increasingly large arrays of random integers and calculated the time required to swap a constant large number ($10^{11}$) of randomly chosen pairs of integers. The results are shown in Fig.~\ref{swap}. Ideally, a swap operation should be $O(1)$ independent of the array size. In practice, it can be seen that larger data sets lead to lower performance as CPU cache is exceeded. The data in Fig.~\ref{swap} was generated by 16 core Intel Xeon 3.33GHz CPUs with 12MB of cache. Our matching code is more complex than this simple swap demonstration, with gradual delocalization of data as the data set increases in size. This leads to a gradual reduction of the probability of a single memory page load containing additional useful data.

\section{Conclusion}
\label{Conclusion}

After accounting for low distance nearby boundaries which limit the complexity of matching (making matching significantly faster) and high distance slower memory access (leading to a slight reduction in performance), Fig.~\ref{timing} provide strong evidence supporting the claimed $O(d^2)$ runtime of our implementation of the algorithm described in \cite{Fowl11b}. A major future goal is to parallelize the algorithm and demonstrate an average processing time per round of error correction independent of the code size using constant computing resources per unit area.

\section{Acknowledgements}

We acknowledge support from the Australian Research Council Centre
of Excellence for Quantum Computation and Communication Technology
(Project number CE110001027), and the US National Security Agency
(NSA) and the Army Research Office (ARO) under contract number
W911NF-08-1-0527.

\section{Supplementary Material}
\label{raw}

The raw data used to generate Figs.~10--11, including timing information, is listed below. The first number is the number of different distances (14). The second is the distance $d$ of the following block of data, the third is the number of different values $n$ of the depolarizing error rate $p$. The next $n$ lines list the value of $p$, $t_{\rm check}$, the number of checks for $Z_L$ state changes, the number of checks for $X_L$ state changes, the observed number of $Z_L$ state changes, the observed number of $X_L$ state changes and finally the total number of CPU seconds devoted to the simulation of that $(d,p)$ pair. This basic structure is repeated for each distance. The last number in the file is the total number of CPU seconds devoted to creation of the entire file. Raw data was obtained using Quad-Core AMD Opteron 2376 processors with 512Kb of cache. Data has also been provided as simple text files.

\footnotesize
\begin{verbatim}
14
3
15
0.0001 25780 93997 93997 11754 10000 37685.82
0.001 261 94887 94886 11810 10000 415.42
0.002 70 93413 93413 12015 10000 228.69
0.005 12 91697 91696 11540 10000 33.62
0.0065 7 97198 97197 12015 10000 30.32
0.007 5 113854 113853 11598 10000 25.31
0.0075 4 121578 121577 11670 10000 24.84
0.008 3 142693 142692 11681 10000 26.08
0.0085 2 184890 184889 11830 10000 29.62
0.009 2 164349 164348 11696 10000 26.93
0.0095 3 107513 107513 11496 10000 28.59
0.01 3 98958 98957 11523 10000 21.47
0.012 2 103971 103970 11759 10000 22.25
0.015 1 125634 125633 11536 10000 27.58
0.05 1 31051 31050 11478 10000 28.47
5
14
0.001 3861 106980 106979 13628 10000 47490.43
0.002 314 151251 151251 13637 10000 6568.45
0.005 26 111948 111947 12953 10000 653.26
0.0065 9 142249 142248 12523 10000 416.50
0.007 6 165703 165702 12626 10000 359.61
0.0075 6 137318 137317 12553 10000 308.21
0.008 6 114309 114308 12021 10000 325.45
0.0085 4 137468 137467 12061 10000 329.63
0.009 3 151197 151196 12121 10000 336.85
0.0095 3 131364 131364 11914 10000 304.38
0.01 3 116216 116215 11896 10000 337.23
0.012 2 103659 103658 11950 10000 332.92
0.015 1 101838 101837 11265 10000 269.22
0.05 1 23807 23806 10975 10000 544.14
7
14
0.001 65493 124652 124652 16178 10000 3521570.27
0.002 3731 114602 114602 15837 10001 165464.09
0.005 88 89778 89777 13864 10000 6818.24
0.0065 13 178531 178530 13492 10000 3272.27
0.007 18 102587 102586 13184 10000 1624.85
0.0075 10 131935 131934 13115 10000 1651.60
0.008 6 157820 157819 12635 10000 1387.70
0.0085 5 148477 148476 12683 10000 1220.78
0.009 4 146237 146236 12497 10000 1470.92
0.0095 4 120724 120724 12117 10000 2172.96
0.01 3 126889 126888 12183 10000 1086.01
0.012 1 157182 157181 11622 10000 1510.58
0.015 1 82346 82345 11007 10000 1145.62
0.05 1 21522 21521 10388 10000 1717.54
9
13
0.002 31862 116333 116333 17719 10001 4289126.50
0.005 134 134774 134773 14912 10000 34427.08
0.0065 31 135372 135371 14001 10000 9139.10
0.007 17 159583 159582 13532 10000 6835.25
0.0075 13 146483 146482 13431 10000 5371.73
0.008 6 206359 206358 12936 10000 4678.65
0.0085 6 157382 157381 12947 10000 3371.82
0.009 6 122371 122370 12735 10000 2893.09
0.0095 3 163207 163207 12108 10000 2943.78
0.01 2 177919 177918 12051 10000 5089.92
0.012 1 140234 140233 11265 10000 4300.04
0.015 1 68981 68980 11432 10000 3114.11
0.05 1 20548 20547 10181 10000 5762.83
11
13
0.002 10000 2970715 2970715 22288 10014 60628087.09
0.005 442 102391 102390 16041 10000 116498.16
0.0065 39 177762 177761 14738 10000 27844.55
0.007 34 126474 126473 13882 10000 26712.36
0.0075 18 149239 149238 13749 10000 15172.28
0.008 10 170139 170138 13115 10000 9234.01
0.0085 8 144574 144573 12947 10000 10480.43
0.009 3 237671 237670 12947 10000 9925.01
0.0095 3 171695 171695 12202 10000 9361.08
0.01 4 107858 107857 11928 10000 6497.89
0.012 1 126752 126751 11515 10000 7390.47
0.015 1 58162 58161 11174 10000 4262.45
0.05 1 19984 19983 9936 10000 12520.50
13
12
0.005 937 114572 114572 16980 10000 603790.28
0.0065 58 192423 192422 15266 10000 78381.30
0.007 32 193240 193239 14502 10000 50053.45
0.0075 26 143566 143565 14055 10000 34005.52
0.008 12 180949 180948 13613 10000 22861.61
0.0085 10 139111 139110 13025 10000 21325.58
0.009 5 172414 172413 12920 10000 20014.77
0.0095 5 121057 121057 12400 10000 13410.58
0.01 3 130958 130957 12065 10000 11380.03
0.012 1 115078 115077 11931 10000 10931.53
0.015 1 50282 50281 11036 10000 6670.03
0.05 1 19964 19963 10009 10000 28043.39
15
12
0.005 2456 102892 102890 17492 10001 2173892.01
0.0065 151 124143 124143 15471 10000 222593.70
0.007 52 176282 176282 15132 10000 115430.02
0.0075 28 174215 174215 14415 10000 78240.36
0.008 17 161912 161912 14015 10000 52890.79
0.0085 11 148576 148576 13102 10000 54511.94
0.009 5 181422 181422 12771 10000 34858.01
0.0095 4 146241 146241 12416 10000 30968.27
0.01 3 129322 129322 12253 10000 28182.51
0.012 1 101864 101864 11498 10000 17875.60
0.015 1 44247 44247 10863 10000 11634.90
0.05 1 19864 19864 9903 10000 98384.02
17
12
0.005 5885 101613 101613 18732 10000 6704496.11
0.0065 275 111939 111939 16021 10000 516700.36
0.007 89 149029 149029 15277 10000 256345.82
0.0075 54 123138 123138 14291 10000 126590.64
0.008 23 148027 148027 13961 10000 82716.06
0.0085 15 127265 127265 13161 10000 56207.63
0.009 5 191569 191569 12961 10000 45131.54
0.0095 4 148226 148226 12348 10000 34119.88
0.01 2 169382 169382 12010 10000 46805.81
0.012 1 89938 89938 11408 10000 24677.61
0.015 1 39250 39250 10892 10000 17671.96
0.05 1 20013 20013 9818 10000 132420.21
19
10
0.0065 420 115409 115409 16474 10000 963439.12
0.007 163 121377 121377 15659 10000 533590.20
0.0075 69 125702 125702 14879 10000 269356.52
0.008 46 94467 94467 13698 10000 144521.98
0.0085 20 114779 114779 13936 10000 82840.47
0.009 8 142942 142942 12913 10000 60109.55
0.0095 4 146009 146009 12300 10000 55487.99
0.01 3 123422 123422 12369 10000 47540.48
0.012 1 81072 81072 11411 10000 35343.85
0.015 1 36705 36705 11159 10000 28088.66
21
10
0.0065 859 91469 91463 16689 10000 2147913.54
0.007 226 124800 124798 16152 10000 1191853.64
0.0075 107 108172 108172 15048 10000 472320.02
0.008 44 118209 118209 14480 10000 276634.69
0.0085 19 128841 128841 13654 10000 125697.36
0.009 8 144541 144541 12716 10000 93677.72
0.0095 3 182294 182294 12630 10000 170367.06
0.01 3 115647 115647 11821 10000 63634.76
0.012 1 75689 75689 11377 10000 52612.18
0.015 1 33355 33355 11109 10000 40737.04
25
5
0.008 55 128564 128564 14352 10000 507860.33
0.0085 17 165192 165192 13888 10000 231570.51
0.009 9 143332 143332 13211 10000 175110.91
0.0095 3 173140 173140 12352 10001 163829.39
0.01 1 9 9 0 1 1.04
35
5
0.008 88 167587 167587 15592 10000 2801803.43
0.0085 14 269115 269115 14131 10000 1292528.08
0.009 7 176942 176942 12822 10000 550365.22
0.0095 4 125106 125106 12314 10001 437366.43
0.01 2 34 34 3 1 61.08
45
5
0.008 230 128455 128422 15969 10000 11334092.96
0.0085 27 203196 203196 14039 10000 2845252.53
0.009 9 147701 147701 12962 10000 1587903.33
0.0095 3 125196 125196 11684 10001 944033.23
0.01 1 13 13 0 1 16.30
55
5
0.008 197 285003 284944 18433 10489 37296797.55
0.0085 34 210799 210785 14340 10000 7669758.24
0.009 7 170072 170072 12541 10000 3025099.15
0.0095 2 140843 140840 11836 10000 2813159.55
0.01 1 66 66 9 1 2038.71
161469013.53
\end{verbatim}
\normalsize

For convenience, we also include below the processed raw data with checks and changes converted into probabilities of logical error. Second column is the probability of $Z_L$ error per round of error correction. Third column is the same data for $X_L$.

\footnotesize
\begin{verbatim}
14
3
15
1.000000e-04 5.581956e-06 4.639970e-06
1.000000e-03 5.480776e-04 4.532594e-04
2.000000e-03 2.119693e-03 1.717972e-03
5.000000e-03 1.193646e-02 1.014745e-02
6.500000e-03 1.987923e-02 1.618780e-02
7.000000e-03 2.227094e-02 1.894943e-02
7.500000e-03 2.594789e-02 2.196915e-02
8.000000e-03 2.892852e-02 2.454551e-02
8.500000e-03 3.308666e-02 2.781694e-02
9.000000e-03 3.694796e-02 3.140969e-02
9.500000e-03 3.853583e-02 3.315384e-02
1.000000e-02 4.229061e-02 3.624911e-02
1.200000e-02 6.016994e-02 5.065678e-02
1.500000e-02 9.182239e-02 7.959676e-02
5.000000e-02 3.696508e-01 3.220606e-01
5
14
1.000000e-03 3.808086e-05 2.680139e-05
2.000000e-03 3.165315e-04 2.257910e-04
5.000000e-03 5.035972e-03 3.770583e-03
6.500000e-03 1.064455e-02 8.347346e-03
7.000000e-03 1.359007e-02 1.060487e-02
7.500000e-03 1.654580e-02 1.294715e-02
8.000000e-03 1.929566e-02 1.577339e-02
8.500000e-03 2.354591e-02 1.927205e-02
9.000000e-03 2.829314e-02 2.309704e-02
9.500000e-03 3.226927e-02 2.678389e-02
1.000000e-02 3.675606e-02 3.050584e-02
1.200000e-02 6.141257e-02 5.081799e-02
1.500000e-02 1.106172e-01 9.819651e-02
5.000000e-02 4.610004e-01 4.200611e-01
7
14
1.000000e-03 2.294325e-06 1.335142e-06
2.000000e-03 4.335013e-05 2.570402e-05
5.000000e-03 2.094460e-03 1.429906e-03
6.500000e-03 6.262974e-03 4.548905e-03
7.000000e-03 8.185048e-03 5.987798e-03
7.500000e-03 1.096092e-02 8.152393e-03
8.000000e-03 1.433186e-02 1.116570e-02
8.500000e-03 1.838769e-02 1.426086e-02
9.000000e-03 2.288849e-02 1.804973e-02
9.500000e-03 2.723837e-02 2.213544e-02
1.000000e-02 3.430413e-02 2.778531e-02
1.200000e-02 7.393956e-02 6.362081e-02
1.500000e-02 1.336675e-01 1.214404e-01
5.000000e-02 4.826679e-01 4.646626e-01
9
13
2.000000e-03 5.701198e-06 2.960670e-06
5.000000e-03 9.323964e-04 5.990253e-04
6.500000e-03 3.723894e-03 2.571852e-03
7.000000e-03 5.436050e-03 3.922925e-03
7.500000e-03 7.731204e-03 5.614499e-03
8.000000e-03 1.103955e-02 8.423489e-03
8.500000e-03 1.475789e-02 1.119869e-02
9.000000e-03 1.907401e-02 1.465237e-02
9.500000e-03 2.606438e-02 2.132012e-02
1.000000e-02 3.509833e-02 2.894034e-02
1.200000e-02 8.033013e-02 7.130980e-02
1.500000e-02 1.657271e-01 1.449695e-01
5.000000e-02 4.954739e-01 4.866886e-01
11
13
2.000000e-03 7.559446e-07 3.382315e-07
5.000000e-03 4.250431e-04 2.457825e-04
6.500000e-03 2.319008e-03 1.527896e-03
7.000000e-03 3.631596e-03 2.525026e-03
7.500000e-03 5.625133e-03 3.980921e-03
8.000000e-03 8.302013e-03 6.213838e-03
8.500000e-03 1.218413e-02 9.219708e-03
9.000000e-03 1.886069e-02 1.443793e-02
9.500000e-03 2.490968e-02 2.022095e-02
1.000000e-02 3.029043e-02 2.499077e-02
1.200000e-02 9.084678e-02 7.889485e-02
1.500000e-02 1.921186e-01 1.719365e-01
5.000000e-02 4.971981e-01 4.999981e-01
13
12
5.000000e-03 1.875611e-04 1.023601e-04
6.500000e-03 1.487202e-03 9.451868e-04
7.000000e-03 2.534618e-03 1.704212e-03
7.500000e-03 4.172993e-03 2.876705e-03
8.000000e-03 6.748340e-03 4.856594e-03
8.500000e-03 1.026050e-02 7.700932e-03
9.000000e-03 1.597595e-02 1.217906e-02
9.500000e-03 2.240651e-02 1.773543e-02
1.000000e-02 3.281617e-02 2.687198e-02
1.200000e-02 1.036773e-01 8.689857e-02
1.500000e-02 2.194815e-01 1.988821e-01
5.000000e-02 4.999981e-01 4.999981e-01
15
12
5.000000e-03 8.458677e-05 4.400683e-05
6.500000e-03 9.483575e-04 5.813465e-04
7.000000e-03 1.807830e-03 1.156577e-03
7.500000e-03 3.220028e-03 2.172807e-03
8.000000e-03 5.559875e-03 3.862835e-03
8.500000e-03 8.742246e-03 6.528671e-03
9.000000e-03 1.494599e-02 1.154505e-02
9.500000e-03 2.272859e-02 1.804911e-02
1.000000e-02 3.381841e-02 2.723163e-02
1.200000e-02 1.128764e-01 9.816980e-02
1.500000e-02 2.455072e-01 2.260046e-01
5.000000e-02 4.985409e-01 4.999981e-01
17
12
5.000000e-03 3.907789e-05 1.862184e-05
6.500000e-03 6.127439e-04 3.577438e-04
7.000000e-03 1.287325e-03 8.089195e-04
7.500000e-03 2.439529e-03 1.638397e-03
8.000000e-03 4.523496e-03 3.145585e-03
8.500000e-03 7.664518e-03 5.666593e-03
9.000000e-03 1.432955e-02 1.090556e-02
9.500000e-03 2.227061e-02 1.779357e-02
1.000000e-02 3.680715e-02 3.044608e-02
1.200000e-02 1.268430e-01 1.111875e-01
1.500000e-02 2.775030e-01 2.547770e-01
5.000000e-02 4.905796e-01 4.996738e-01
19
10
6.500000e-03 4.000280e-04 2.265075e-04
7.000000e-03 9.146123e-04 5.520112e-04
7.500000e-03 1.953781e-03 1.254145e-03
8.000000e-03 3.708987e-03 2.579145e-03
8.500000e-03 6.906109e-03 4.763698e-03
9.000000e-03 1.230079e-02 9.332162e-03
9.500000e-03 2.253916e-02 1.807942e-02
1.000000e-02 3.592503e-02 2.861376e-02
1.200000e-02 1.407514e-01 1.233468e-01
1.500000e-02 3.040190e-01 2.724428e-01
21
10
6.500000e-03 2.641856e-04 1.436088e-04
7.000000e-03 6.622730e-04 3.862715e-04
7.500000e-03 1.521229e-03 9.543867e-04
8.000000e-03 3.183285e-03 2.101901e-03
8.500000e-03 6.229161e-03 4.419605e-03
9.000000e-03 1.195018e-02 9.221913e-03
9.500000e-03 2.425194e-02 1.899817e-02
1.000000e-02 3.669981e-02 3.066557e-02
1.200000e-02 1.503129e-01 1.321197e-01
1.500000e-02 3.330526e-01 2.998056e-01
25
5
8.000000e-03 2.291625e-03 1.534796e-03
8.500000e-03 5.385383e-03 3.781245e-03
9.000000e-03 1.119275e-02 8.279782e-03
9.500000e-03 2.501067e-02 2.004721e-02
1.000000e-02 0.000000e+00 1.111112e-01
35
5
8.000000e-03 1.168454e-03 7.215547e-04
8.500000e-03 3.946915e-03 2.750413e-03
9.000000e-03 1.105951e-02 8.494526e-03
9.500000e-03 2.666542e-02 2.130889e-02
1.000000e-02 4.625748e-02 1.492872e-02
45
5
8.000000e-03 6.210500e-04 3.678873e-04
8.500000e-03 2.746342e-03 1.915113e-03
9.000000e-03 1.060803e-02 8.018221e-03
9.500000e-03 3.327366e-02 2.818675e-02
1.000000e-02 0.000000e+00 7.692313e-02
55
5
8.000000e-03 3.514479e-04 1.940539e-04
8.500000e-03 2.146034e-03 1.463901e-03
9.000000e-03 1.126806e-02 8.856805e-03
9.500000e-03 4.394984e-02 3.685991e-02
1.000000e-02 1.363635e-01 1.515150e-02
\end{verbatim}
\normalsize

\bibliography{../References}

\begin{thebibliography}{17}
\expandafter\ifx\csname natexlab\endcsname\relax\def\natexlab#1{#1}\fi
\expandafter\ifx\csname bibnamefont\endcsname\relax
  \def\bibnamefont#1{#1}\fi
\expandafter\ifx\csname bibfnamefont\endcsname\relax
  \def\bibfnamefont#1{#1}\fi
\expandafter\ifx\csname citenamefont\endcsname\relax
  \def\citenamefont#1{#1}\fi
\expandafter\ifx\csname url\endcsname\relax
  \def\url#1{\texttt{#1}}\fi
\expandafter\ifx\csname urlprefix\endcsname\relax\def\urlprefix{URL }\fi
\providecommand{\bibinfo}[2]{#2}
\providecommand{\eprint}[2][]{\url{#2}}

\bibitem[{\citenamefont{Shor}(1994)}]{Shor94b}
\bibinfo{author}{\bibfnamefont{P.~W.} \bibnamefont{Shor}}, in
  \emph{\bibinfo{booktitle}{Proc. 35th Annual Symposium on Foundations of
  Computer Science}} (\bibinfo{publisher}{IEEE Computer Society Press},
  \bibinfo{address}{Los Alamitos, CA}, \bibinfo{year}{1994}), pp.
  \bibinfo{pages}{124--134}, \bibinfo{note}{quant-ph/9508027}.

\bibitem[{\citenamefont{Lloyd}(1996)}]{Lloy96}
\bibinfo{author}{\bibfnamefont{S.}~\bibnamefont{Lloyd}},
  \bibinfo{journal}{Science} \textbf{\bibinfo{volume}{273}},
  \bibinfo{pages}{1073} (\bibinfo{year}{1996}).

\bibitem[{\citenamefont{Jordan}(2012)}]{Jord10}
\bibinfo{author}{\bibfnamefont{S.}~\bibnamefont{Jordan}},
  \emph{\bibinfo{title}{Quantum algorithm zoo}},
  \bibinfo{howpublished}{http://math.nist.gov/quantum/zoo/}
  (\bibinfo{year}{2012}).

\bibitem[{\citenamefont{Wang et~al.}(2011)\citenamefont{Wang, Fowler, and
  Hollenberg}}]{Wang11}
\bibinfo{author}{\bibfnamefont{D.~S.} \bibnamefont{Wang}},
  \bibinfo{author}{\bibfnamefont{A.~G.} \bibnamefont{Fowler}},
  \bibnamefont{and} \bibinfo{author}{\bibfnamefont{L.~C.~L.}
  \bibnamefont{Hollenberg}}, \bibinfo{journal}{Phys. Rev. A}
  \textbf{\bibinfo{volume}{83}}, \bibinfo{pages}{020302(R)}
  (\bibinfo{year}{2011}), \bibinfo{note}{arXiv:1009.3686}.

\bibitem[{\citenamefont{Fowler et~al.}(2012{\natexlab{a}})\citenamefont{Fowler,
  Whiteside, and Hollenberg}}]{Fowl11b}
\bibinfo{author}{\bibfnamefont{A.~G.} \bibnamefont{Fowler}},
  \bibinfo{author}{\bibfnamefont{A.~C.} \bibnamefont{Whiteside}},
  \bibnamefont{and} \bibinfo{author}{\bibfnamefont{L.~C.~L.}
  \bibnamefont{Hollenberg}}, \bibinfo{journal}{Phys. Rev. Lett.}
  \textbf{\bibinfo{volume}{108}}, \bibinfo{pages}{180501}
  (\bibinfo{year}{2012}{\natexlab{a}}), \bibinfo{note}{arXiv:1110.5133}.

\bibitem[{\citenamefont{Bravyi and Kitaev}(1998)}]{Brav98}
\bibinfo{author}{\bibfnamefont{S.~B.} \bibnamefont{Bravyi}} \bibnamefont{and}
  \bibinfo{author}{\bibfnamefont{A.~Y.} \bibnamefont{Kitaev}},
  \bibinfo{journal}{quant-ph/9811052}  (\bibinfo{year}{1998}).

\bibitem[{\citenamefont{Dennis et~al.}(2002)\citenamefont{Dennis, Kitaev,
  Landahl, and Preskill}}]{Denn02}
\bibinfo{author}{\bibfnamefont{E.}~\bibnamefont{Dennis}},
  \bibinfo{author}{\bibfnamefont{A.}~\bibnamefont{Kitaev}},
  \bibinfo{author}{\bibfnamefont{A.}~\bibnamefont{Landahl}}, \bibnamefont{and}
  \bibinfo{author}{\bibfnamefont{J.}~\bibnamefont{Preskill}},
  \bibinfo{journal}{J. Math. Phys.} \textbf{\bibinfo{volume}{43}},
  \bibinfo{pages}{4452} (\bibinfo{year}{2002}),
  \bibinfo{note}{quant-ph/0110143}.

\bibitem[{\citenamefont{Raussendorf and Harrington}(2007)}]{Raus07}
\bibinfo{author}{\bibfnamefont{R.}~\bibnamefont{Raussendorf}} \bibnamefont{and}
  \bibinfo{author}{\bibfnamefont{J.}~\bibnamefont{Harrington}},
  \bibinfo{journal}{Phys. Rev. Lett.} \textbf{\bibinfo{volume}{98}},
  \bibinfo{pages}{190504} (\bibinfo{year}{2007}),
  \bibinfo{note}{quant-ph/0610082}.

\bibitem[{\citenamefont{Raussendorf et~al.}(2007)\citenamefont{Raussendorf,
  Harrington, and Goyal}}]{Raus07d}
\bibinfo{author}{\bibfnamefont{R.}~\bibnamefont{Raussendorf}},
  \bibinfo{author}{\bibfnamefont{J.}~\bibnamefont{Harrington}},
  \bibnamefont{and} \bibinfo{author}{\bibfnamefont{K.}~\bibnamefont{Goyal}},
  \bibinfo{journal}{New J. Phys.} \textbf{\bibinfo{volume}{9}},
  \bibinfo{pages}{199} (\bibinfo{year}{2007}),
  \bibinfo{note}{quant-ph/0703143}.

\bibitem[{\citenamefont{Fowler et~al.}(2012{\natexlab{b}})\citenamefont{Fowler,
  Mariantoni, Martinis, and Cleland}}]{Fowl12f}
\bibinfo{author}{\bibfnamefont{A.~G.} \bibnamefont{Fowler}},
  \bibinfo{author}{\bibfnamefont{M.}~\bibnamefont{Mariantoni}},
  \bibinfo{author}{\bibfnamefont{J.~M.} \bibnamefont{Martinis}},
  \bibnamefont{and} \bibinfo{author}{\bibfnamefont{A.~N.}
  \bibnamefont{Cleland}}, \bibinfo{journal}{Phys. Rev. A}
  \textbf{\bibinfo{volume}{86}}, \bibinfo{pages}{032324}
  (\bibinfo{year}{2012}{\natexlab{b}}), \bibinfo{note}{arXiv:1208.0928}.

\bibitem[{\citenamefont{Fowler et~al.}(2012{\natexlab{c}})\citenamefont{Fowler,
  Whiteside, McInnes, and Rabbani}}]{Fowl12d}
\bibinfo{author}{\bibfnamefont{A.~G.} \bibnamefont{Fowler}},
  \bibinfo{author}{\bibfnamefont{A.~C.} \bibnamefont{Whiteside}},
  \bibinfo{author}{\bibfnamefont{A.~L.} \bibnamefont{McInnes}},
  \bibnamefont{and} \bibinfo{author}{\bibfnamefont{A.}~\bibnamefont{Rabbani}},
  \bibinfo{journal}{arXiv:1202.6111}  (\bibinfo{year}{2012}{\natexlab{c}}).

\bibitem[{\citenamefont{Edmonds}(1965{\natexlab{a}})}]{Edmo65a}
\bibinfo{author}{\bibfnamefont{J.}~\bibnamefont{Edmonds}},
  \bibinfo{journal}{Canad. J. Math.} \textbf{\bibinfo{volume}{17}},
  \bibinfo{pages}{449} (\bibinfo{year}{1965}{\natexlab{a}}).

\bibitem[{\citenamefont{Edmonds}(1965{\natexlab{b}})}]{Edmo65b}
\bibinfo{author}{\bibfnamefont{J.}~\bibnamefont{Edmonds}}, \bibinfo{journal}{J.
  Res. Nat. Bur. Standards} \textbf{\bibinfo{volume}{69B}},
  \bibinfo{pages}{125} (\bibinfo{year}{1965}{\natexlab{b}}).

\bibitem[{\citenamefont{Gottesman}(1997)}]{Gott97}
\bibinfo{author}{\bibfnamefont{D.}~\bibnamefont{Gottesman}}, Ph.D. thesis,
  \bibinfo{school}{Caltech} (\bibinfo{year}{1997}),
  \bibinfo{note}{quant-ph/9705052}.

\bibitem[{\citenamefont{Duclos-Cianci and Poulin}(2010)}]{Ducl09}
\bibinfo{author}{\bibfnamefont{G.}~\bibnamefont{Duclos-Cianci}}
  \bibnamefont{and} \bibinfo{author}{\bibfnamefont{D.}~\bibnamefont{Poulin}},
  \bibinfo{journal}{Phys. Rev. Lett.} \textbf{\bibinfo{volume}{104}},
  \bibinfo{pages}{050504} (\bibinfo{year}{2010}),
  \bibinfo{note}{arXiv:0911.0581}.

\bibitem[{\citenamefont{Wootton and Loss}(2012)}]{Woot12}
\bibinfo{author}{\bibfnamefont{J.~R.} \bibnamefont{Wootton}} \bibnamefont{and}
  \bibinfo{author}{\bibfnamefont{D.}~\bibnamefont{Loss}},
  \bibinfo{journal}{arXiv:1202.4316}  (\bibinfo{year}{2012}).

\bibitem[{\citenamefont{Bravyi et~al.}(2012)\citenamefont{Bravyi,
  Duclos-Cianci, Poulin, and Suchara}}]{Brav12}
\bibinfo{author}{\bibfnamefont{S.}~\bibnamefont{Bravyi}},
  \bibinfo{author}{\bibfnamefont{G.}~\bibnamefont{Duclos-Cianci}},
  \bibinfo{author}{\bibfnamefont{D.}~\bibnamefont{Poulin}}, \bibnamefont{and}
  \bibinfo{author}{\bibfnamefont{M.}~\bibnamefont{Suchara}},
  \bibinfo{journal}{arXiv:1207.1443}  (\bibinfo{year}{2012}).

\end{thebibliography}

\end{document}